# Approaching the quantum-limited precision in frequency-comb-based spectral interferometry for length measurements

Yoon-Soo Jang[1,2,*], Heulbi Ahn[3], Sunghoon Eom[1], Jungjae Park[1,2], and Jonghan Jin[1,2]

[1]Length Standard Group, Division of Physical Metrology, Korea Research Institute of Standards and Science (KRISS), 267 Gajeong-ro, Yuseong-gu, Daejeon, 34113, Republic of Korea

[2] Department of Science of Measurement, University of Science and Technology (UST), Daejeon, 34113, Republic of Korea

[3]Meterlab Corp., Daejeon, 34113, Republic of Korea

*ysj@kirss.ac.kr

**Abstract**

**Over the last two decades, frequency combs have brought breakthroughs in length metrology with traceability to length standards. In particular, frequency-comb-based spectral interferometry is regarded as a promising technology for next-generation length standards. However, to achieve this, the nanometer-level precision inherent in laser interferometer is required. Here, we report distance measurements by a frequency-comb-based spectral interferometry with sub-nm precision close to a standard quantum limit. The measurement precision was confirmed as 0.67 nm at an averaging time of 25 μs. The measurement sensitivity was found to be $4.5{\cdot}10^{-12}$m/Hz$^{1/2}$, close to the quantum-limit. As a practical example of observing precise physical phenomena, we demonstrated measurements of acoustic-wave-induced vibration and laser eavesdropping. Our study will be an important step toward the practical realization of upcoming length standards.**

## Main text

### Introduction

Length is one of the seven fundamental physical quantities in the SI unit system, and the capability of its precise and accurate measurement it is essential not only for science and technology but also for everyday life (*1-3*). The most precise and accurate way to measure a physical quantity is to measure based on the definition of that quantity. The current definition of the length unit "*meter*" is defined as follows: “The metre, symbol m, is defined by taking the fixed numerical value of the speed of light in vacuum *c* to be 299 792 458 when expressed in the unit m $s^{-1}$” (*4*). The *mises en pratique* regarding the definition of the meter involves an interferometric distance measurement based on well-defined wavelengths recommended by the Bureau International des poids et mesures (BIPM) (*5*). Displacement interferometers based on single-wavelength lasers are the most common method for measuring length according to the definition of the length unit "m". This method allows for sub-nanometer level resolutions, but due to the 2π ambiguity problem, it requires the accumulated displacement from the initial position to measure the length. Therefore, stable environmental conditions are necessary during measurements, and there are many constraints on its use.

To overcome these limitations, various technologies for absolute distance measurements have been demonstrated (*6*). However, the measurement precision of such methods is at the micrometer level, which is not enough to replace the laser displacement interferometry in the ultra-precision measurement field. Spectral interferometry is an interferometric technique capable of measuring absolute distances at the sub-femtosecond level resolution in the time domain and nanometer-level resolutions in the length domain by analyzing interference patterns, which appear as the reciprocal of the time delay ($1/\tau_{TOF}$) in the frequency domain (*7,8*). However, conventional spectral interferometers using broadband light sources such as light emitting diodes, with their short coherence length, are used in applications where the measurement range is on the order of a few millimeters, such as in OCT and industrial sample thickness measurements (*9,10*).

Recently, the advent of the frequency comb has brought a breakthrough in optical metrology (*11,12*). Frequency combs have a wide spectrum in the frequency domain, similar to low-coherence light sources, while the individual modes of frequency combs are precisely evenly spaced and have very narrow linewidths. Furthermore, frequency combs can be referenced to frequency standards, enabling highly reliable and accurate optical metrology with traceability.

Based on these characteristics, frequency combs have contributed significantly to the advancement of length metrology over the past 20 years (*13*), including dual-comb ranging, (*14,15*) multi-wavelength interferometry (*16,17*), amplitude-modulated continuous-wave ranging (*18*), frequency-modulated continuous-wave ranging (*19,20*), pulse-to-pulse interferometry (*21*), timing-synchronization-based ranging (*22,23*), and spectral interferometry (*24-27*).

For spectral interferometry, these characteristics of frequency combs enable long-distance measurements with sub-μm precision (*28-31*). Furthermore, a high-repetition-rate frequency comb (*32*) enables comb-mode-resolved spectral interferometer for arbitrary distance measurements (*33*). With these advantages, frequency-comb-based spectral interferometry is considered as a promising technology for next-generation length standards, potentially replacing the laser displacement interferometer, which is the current length standard that is also widely used in precision engineering. To enable frequency-comb-based spectral interferometry to replace laser displacement interferometry, nanometer-level measurement precision that laser displacement interferometry offers must be realized. Many techniques, such as the use of broadband light sources and improved data processing algorithms, have been proposed to enhance the measurement precision of spectral interferometry (*34,35*). However, the fundamental precision limits in frequency-comb-based spectral interferometry has yet to be reported.

Here, we present distance measurements using a frequency-comb-based spectral interferometry with sub-nm precision close to a standard quantum limit and a study of the fundamental limit of its measurement precision. In frequency-comb-based spectral interferometry, the factors influencing the measurement precision include the spectral shape, the frequency noise and the intensity noise. In previous studies by the authors, it was experimentally and demonstrated through a simulation that the measurement precision can be improved with a broader bandwidth of the light source (*35*). Based on our theoretical model, we devised a reasonable hypothesis that intrinsic factors influencing the measurement precision in frequency-comb-based spectral interferometry are the intensity noise and the frequency noise. The intensity noise is independent of the target distance and dominant at short distances, considering frequency uncertainty levels of less than $10^{-12}$ for a frequency comb locked to a frequency standard (*36*). Frequency noise is dependent on the target distance and would be dominant at long distances (*17*). The purpose of this study is to validate our hypothesis and explore the origin of the measurement precision of frequency-comb-based spectral interferometry. In this study, a broadband and spectrally flat electro-optic frequency comb (EO comb) and a high-speed spectrometer were used to construct a frequency-comb-based spectral interferometer for ultraprecision absolute distance measurements. To analyze the measurement precision of this interferometer, measurement precision in terms of the Allan deviation was experimentally confirmed to be 0.67 nm (0.34 nm) at an averaging time of 25 μs (250 μs). For a frequency-domain analysis, the amplitude spectral density was observed to be $4.5 \cdot 10^{-12}$m/Hz$^{1/2}$ in the range of > 1 kHz, close to the quantum-limited (shot-noise-limited) value of $3.2 \cdot 10^{-12}$m/Hz$^{1/2}$. To estimate the measurement precision theoretically, we established a modeling equation that includes both the intensity noise and the frequency noise. To investigate the influence of the intensity noise on the measurement precision, we measured the intensity noise of the EO comb using the same spectrometer and injected this noise into the spectral interferogram signal. To investigate the influence of the frequency noise on the measurement precision, we analyzed the measurement sensitivity as a function of the target distance from 100 mm to 1,000 mm. Besides, our capability of ultra-precision and high-speed distance measurements can serve as a new measurement platform for optomechanical motion (*37*), dynamic deformation (*38*), and ultrasonic dynamics (*39*). As an example of the realization of dynamic motion measurements, we measured micro-vibrations generated by sound waves using the capability of ultra-precise and real-time distance measurements.

## Results

### Modeling of spectral interference signals with intensity noise to determine the origin of the precision limits in frequency-comb-based spectral interferometry

In this study, a broadband and spectrally flat custom-made EO comb with a repetition rate ($f_r$) of 18 GHz was employed as a light source. The frequency of the $i$-th frequency comb mode ($f_i$) can be expressed as $f_i = f_{seed} + i \cdot f_r$, where $f_{seed}$ is the frequency of the seed laser and $i$ is a real number ($-N$ … 0 … $N$) (see further details in Appendix C) (*40*). Spectral interference was generated by an all-fiber interferometer and measured by a high-speed spectrometer (s-Nova-1550, METERLAB) (see further details in the Methods section). Spectral interferometry can measure a target distance ($L$) by analyzing the spectral interference generated by the time delay ($\tau_{TOF}=2L/v$), where $v$ is the speed of light in air (*27*). Spectral interference has a period of $1/\tau_{TOF}$ in the frequency domain, as shown in Fig. 1A (see further details in Appendix A).

Assuming that the target distance is fixed and that the intensity levels of the reference and measurement beams are identical to simplify the modeling of the spectral interference, the spectral interference signal ($I(f_i,t)$) of the $i$-th frequency comb mode ($f_i$) can be expressed as

$$I(f_i, t) = I_o(f_i, t) \cdot \{1 + V \cos\left(2\pi(f_i + \delta_{fi}(t)) \cdot 2L/v\right)\} \tag{1}$$

where $I_o(f_i,t)$ is the intensity of the $i$-th frequency comb mode, $V$ is the visibility, $\delta_{fi}(t)$ is the frequency offset from ideal frequency of $f_i$, which acts as frequency noise. In frequency-comb-based spectral interferometry, the frequency of the frequency comb is considered to be fixed. Therefore, intrinsic noise sources such as intensity ($\Delta I_o(f_i, t)$) noise and frequency noise ($\Delta\delta_{fi}(t)$) can be interpreted as affecting the amplitude of the spectral interference and the corresponding Fourier transformation. Fluctuations of the spectral interference can be expressed as shown below:

$$(\Delta I(f_i, t))^2 = \left(\partial I/_{\partial I_o}\right)^2 \cdot (\Delta I_o(f_i, t))^2 + \left(\partial I/_{\partial f_{o,i}}\right)^2 \cdot (\Delta\delta_{fi}(t))^2 \tag{2}$$

Fluctuations of the spectral interference cause fluctuations in the distance measurements. In this study, we simplified the fluctuation of the spectral interference signal as follows:

$$(\Delta I(f_i, t))^2 \approx (1 + V^2/2) \cdot (\Delta I_o(f_i, t))^2 + (4\pi I_o V L/v)^2/2 \cdot (\Delta\delta_{fi}(t))^2 \tag{3}$$

The term on the left and that on the right denote the fluctuation of the spectral interference induced from intensity and frequency changes of the light source, respectively. This model equation was utilized to simulate the measurement precision of frequency-comb-based spectral interferometry influenced by intensity noise and frequency noise. To validate our hypothesis, we compare these simulation results with experimental results.

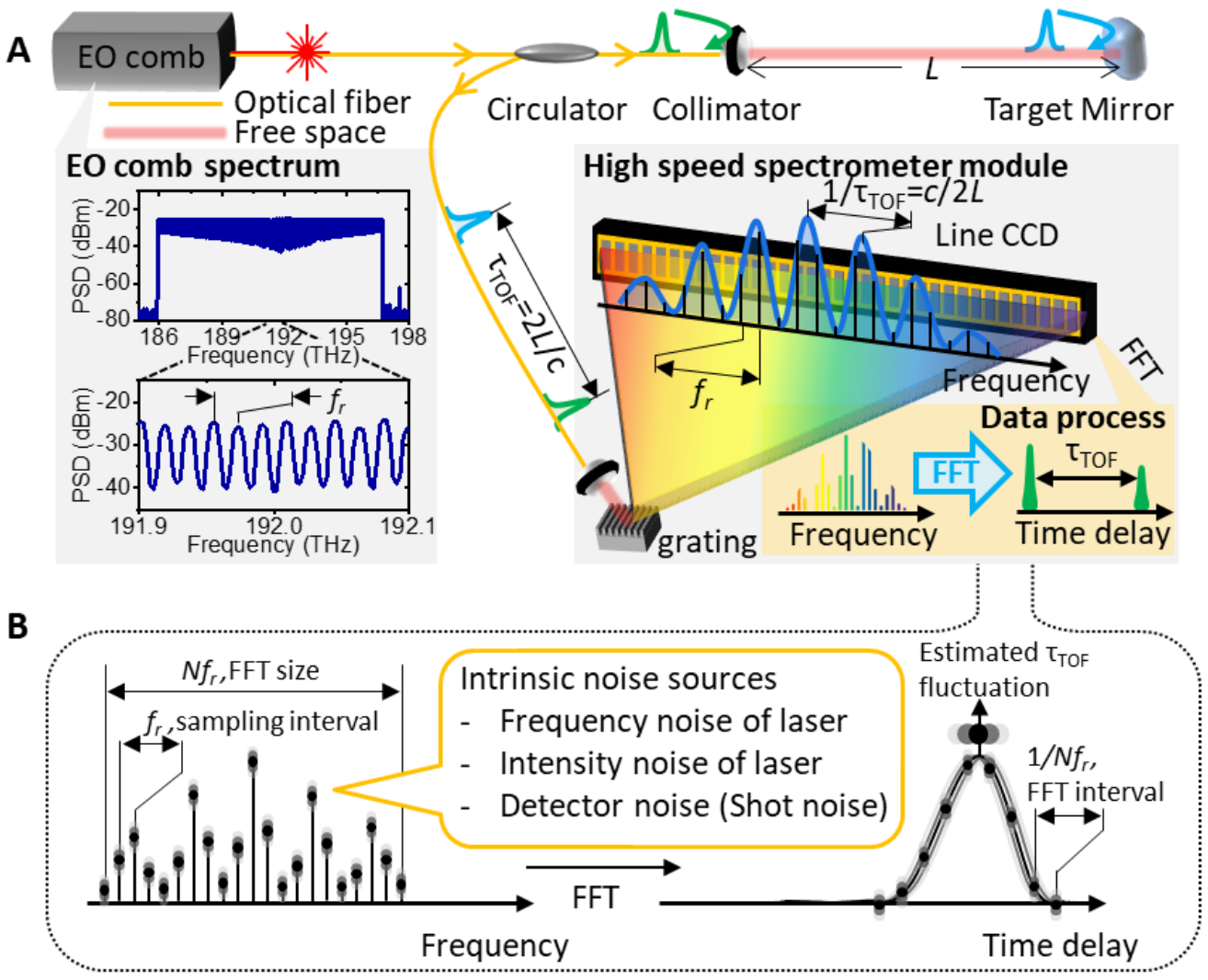

**FIG. 1. Optical layout of EO comb spectral interferometric ranging and conceptual images of the corresponding precision limit: A,** Optical layout of EO-comb-based spectral interferometry for distance measurements. The left inset shows the EO comb spectrum. The right inset shows the high-speed spectrometer module for measuring the interference spectra to determine the absolute distance. **B,** Intrinsic noise sources of the precision limit of frequency comb spectral interferometry. With a frequency comb well defined in the frequency domain, the frequency noise can be negligible. Intensity noise sources including the light source and the detector in the frequency domain induce amplitude noise in the time-domain signal generated through Fourier transformation, which limits the measurement precision.

**High-precision and rapid distance measurements with frequency-comb-based spectral interferometry**

Figure 2 shows the experimental results of distance measurements using frequency-comb-based spectral interferometry and the intensity noise effects on the measurement precision. A target mirror was fixed at a distance of about 100 mm, and distance measurements were conducted for one second at an update rate of 40 kHz. We utilized AMCW ranging to determine the initial value for the distance measurement, considering the target distance beyond the non-ambiguity range of 4.2 mm in our frequency-comb-based spectral interferometry (*41*). As in our previous study, we applied a super-Gaussian window with a 9 THz bandwidth for data processing during the distance measurements (*35*) (see further details in Appendices B and D). To verify the effects of the intensity noise on the measurement precision, the intensity noise of the light source ($\Delta I_o(f_i, t)$) was measured by a high-speed spectrometer and its contribution ($\sqrt{1+V^2/2}\cdot\Delta I_o(f_i,t)$) was added to the single spectral interferogram ($I(f_i,t_o)$) measured by the same high-speed spectrometer at $t=t_o$. Note that the intensity noise takes into account the light source and the high-speed spectrometer.

As shown in Fig. 2A, the mean value of the measured distance (blue line) was 100.015303 mm with a standard deviation (1 σ) of 2.3 nm over 1 s. The standard deviation (1 σ) predicted from the intensity noise (yellow line) was found to be 1.8 nm. For a further analysis of the measurement results in the frequency domain, amplitude spectral density levels from 10 Hz to 20 kHz were calculated based on the time-domain measurement results shown in Fig. 2a. From 1 kHz to 20 kHz, the amplitude spectral density was dominated by white noise at the level of $4.5\cdot10^{-12}$m/Hz$^{1/2}$, which was close to the quantum limit (shot noise limit) ($3.7\cdot10^{-12}$m/Hz$^{1/2}$), as discussed later. Below 2 kHz, the amplitude spectral density was dominated by flicker noise ($1/f^{1/2}$ noise) and random walk noise ($1/f$ noise). In addition, peak components originating from technical noise, such as uncompensated acoustic noise, were observed. For example, the largest peaks near 7.8 kHz were generated from the electrical DC power source. The shot noise level of the spectrometer in terms of the relative intensity noise was $1.7\cdot10^{-11}$/Hz (or -107.7 dBc/Hz), corresponding to an amplitude spectral density of $3.7\cdot10^{-12}$m/Hz$^{1/2}$ (see further details in the Methods section). Both the experimental and predicted results of the amplitude spectral density were $4.5\cdot10^{-12}$m/Hz$^{1/2}$, as shown in Fig. 2b. This discrepancy was attributed to the fact that intensity noise in certain frequency bands slightly exceeded the shot noise limit (see further details in Appendix E). Fig. 2B also shows the amplitude spectral densities from the simulation results, considering only the intensity noise, to validate our hypothesis. We analyzed the influence of the intensity noise on the distance measurements by processing the spectral interferogram with the injected intensity noise. In the frequency domain, the amplitude spectral densities from the experimental results (blue line) and the predicted results (yellow line) with the intensity noise were in good agreement from 10 Hz to 20 kHz, as shown in Fig.2B.

To evaluate the measurement precision, the Allan deviation of the measurement results (blue line) and the prediction from the intensity noise (yellow line) were calculated as shown in Fig. 2C. The Allan deviation value of the measurement results was 0.67 nm at an averaging time of 25 μs (without averaging) and decreased gradually to 0.34 nm at an averaging time of 250 μs, as shown in Fig. 2C. At shorter averaging times of 25 μs to 2.5 ms, the Allan deviations of both the experimental and predicted results were also in good agreement with the white-noise-limited Allan deviation value of 3.2 $\mathrm{pm}\cdot\tau^{-1/2}$ as estimated using an Enrico chart (*42*). It should be noted that the quantum limited (shot noise limit) Allan deviation was calculated and found to be 2.7 $\mathrm{pm}\cdot\tau^{-1/2}$. Over an averaging time of 250 μs, in both cases the Allan deviation increased gradually due to flicker noise and random walk noise, reaching approximately 1 nm at an averaging time of 0.1 s. This outcome can be attributed to actual distance variations caused by thermal expansion in the laboratory environment (*41,43*). From an averaging time of 10 μs to 5 ms, the Allan deviation of the experimental results slightly exceeded the predicted results, likely due to actual distance variations caused by acoustic noise in the laboratory environment. The Allan deviations of both the experimental and predicted results were also in good agreement. Through these results, we confirmed that the measurement precision of our frequency-comb-based spectral interferometer was primarily limited by the intensity noise.

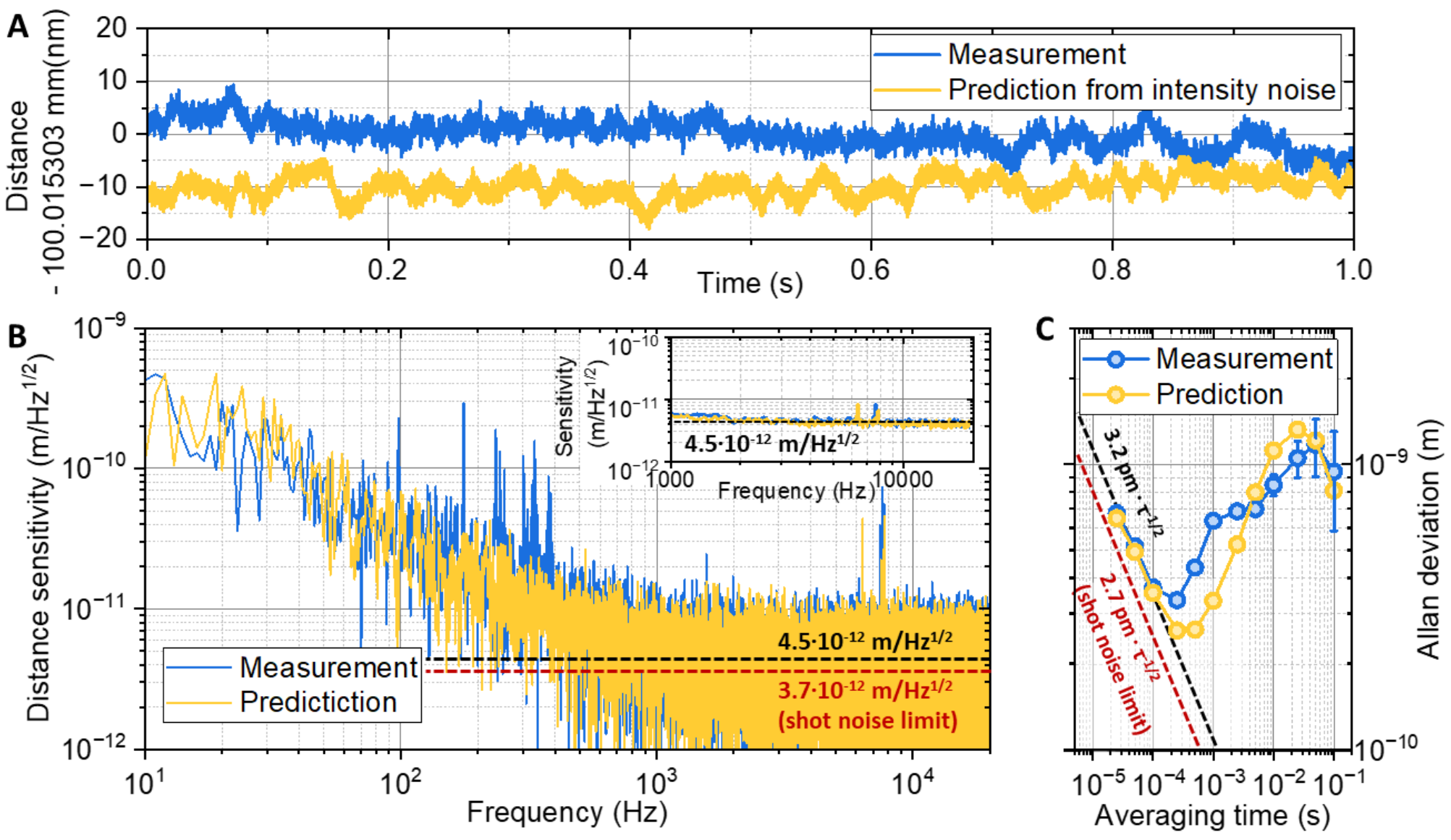

**FIG. 2. Experimental demonstration of EO comb spectral interferometric ranging and comparison with intensity noise-based predictions: A**, Distance measurement (black) with an update rate of 40 kHz over 1 s, corresponding to 40,000 data points. The yellow line shows the prediction from the intensity noise. **B**, Distance sensitivity from 10 Hz to 20 kHz of the Nyquist frequency in an amplitude spectral density unit. The black line shows the measurement results and the yellow line shows the prediction from the intensity noise. The inset shows 100-point moving-average lines of Fig. 2b from 1 kHz to 20 kHz. **C**, Allan deviation from 25 μs to 100 ms to evaluate the measurement precision.

To evaluate the effects of frequency noise on the measurement precision, we analyzed the distance sensitivity according to the target distance from 100 mm to 1,000 mm, as shown in Figs. 3A and B. From 10 Hz to 500 Hz, where $1/f$ noise is dominant, the observed distance sensitivity outcomes were from 100 mm to 500 mm and showed no significant differences. They were also limited by intensity noise, as presented in Fig. 2. The distance sensitivity observed at 700 mm and 1,000 mm appeared slightly higher due to the actual distance variation caused by changes in the surrounding environment. In the region above 1 kHz, we observed that the distance sensitivity increased gradually as the target distance was increased. The white-noise-limited distance sensitivity above 10 kHz observed from 100 mm to 1,000 mm is plotted in Fig. 3c to identify the trend in this case. The intensity noise contributed by $4.5 \cdot 10^{-12}$ m/Hz$^{1/2}$ and frequency noise contributed by $L \cdot 1.2 \cdot 10^{-11}$ m/Hz$^{1/2}$, where $L$ in meter, as estimated using the frequency noise (see further details in the Methods section). The length-dependent white-noise-limited sensitivity ($S_{prediction}(L)$) can be estimated as $S_{prediction}(L) = \sqrt{(4.5 \cdot 10^{-12})^2 + (L \cdot 1.2 \cdot 10^{-11})^2}\ \mathrm{m/Hz^{1/2}}$, as presented by the yellow line of Fig. 3C. Our estimated line and measurement results were in good agreement. Through these results, we confirmed that the measurement precision of our frequency-comb-based spectral interferometer was mainly limited by frequency noise over longer distances.

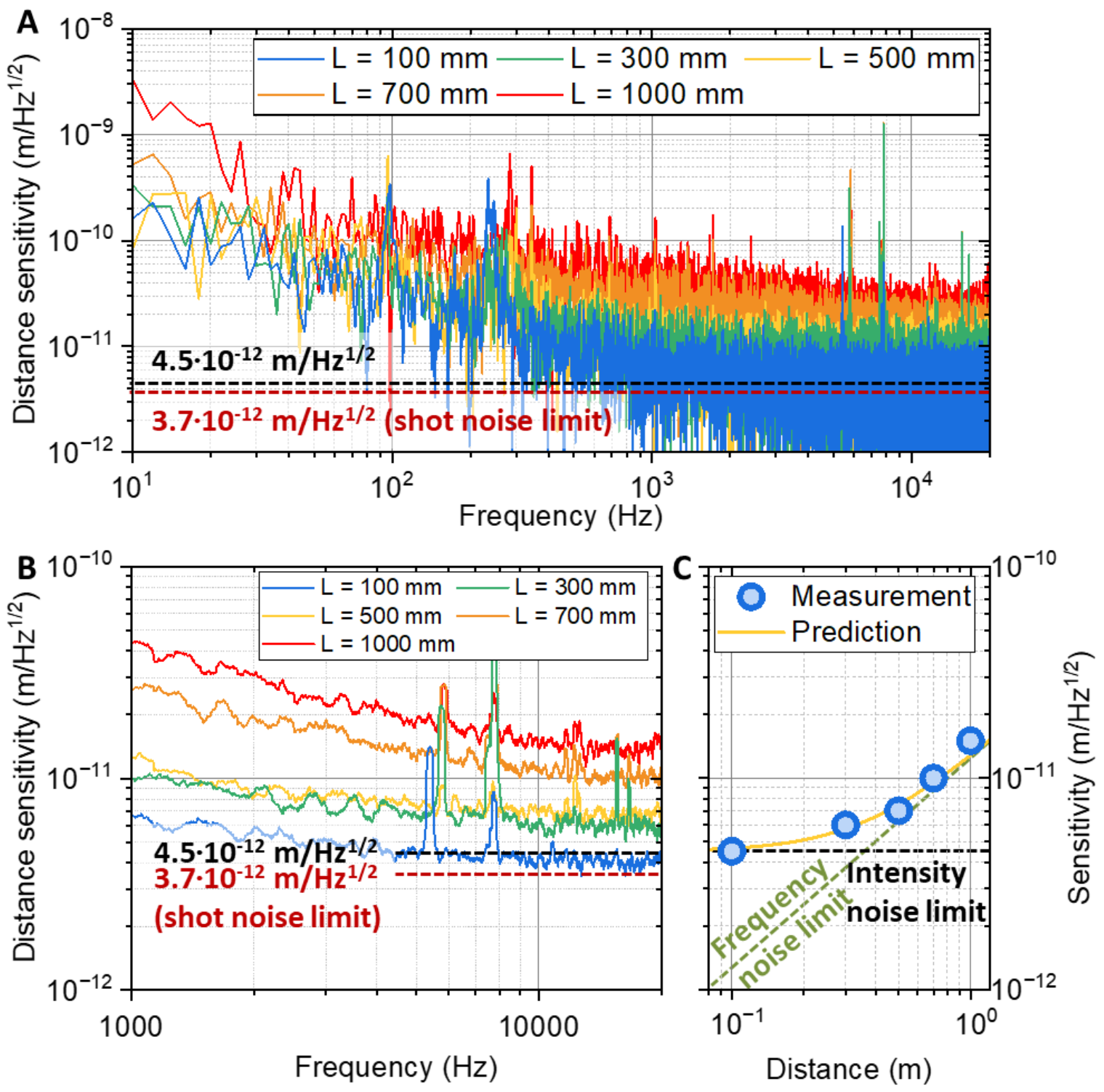

**FIG. 3. Distance-dependent variation of the distance sensitivity to evaluate the frequency noise effects on the measurement precision: A**, Distance sensitivity for target distances from 100 mm to 1,000 mm. **B**, 100-point moving-average lines of Fig. 3a from 10 kHz to 20 kHz. **C**, White noise limit of the distance sensitivity for various distances with a comparison with the intensity noise and frequency noise limits.

**Sound-induced vibration measurements**

The ultra-precise and real-time capabilities of the proposed length measurement method make this method suitable for sound-induced vibration measurements. Fig. 4 shows the measurement results of sound-wave-induced vibrations using the proposed ultra-precise and real-time distance-measurement method. We generated sounds at frequencies of 1 kHz, 3 kHz, 5 kHz and 10 kHz using a mobile phone behind a pellicle beam splitter. Subsequently, we measured sound-wave-induced vibrations on the surface of the pellicle beam splitter. Figs. 4A to D show the vibrations at each frequency plotted for approximately ten periods, indicating that they were clearly observed in both the time and frequency domain.

In addition, we recorded frequency-modulated sound and a human voice via our frequency-comb-based spectral interferometry, as shown in Fig. 4 E-1 and F-1, respectively. Both sound levels were around 70 ~ 80 decibels. The frequency-modulated sound and the human voice were converted into the ".wav" file format (see supplementary multimedia), and these ".wav" files were converted into spectrograms, as shown in Figs. 4 E-2 and F-2, respectively. The frequency-modulated sound has a center frequency of 5 kHz with a modulation depth of 1 kHz and modulation period of 0.5 s. (see also audio S1) The frequency-modulated sound was clearly observed in the time domain, and its spectrogram clearly shows the frequency-modulated sound signal. The human-voice-induced vibration has an amplitude of 200 nm. The words "KRISS length standard" was clearly audible, as shown in Fig. 4 F-1 and the supplementary multimedia (audio S2). As shown in Fig. 4 F-2, the spectrogram of the recorded voice clearly shows the human voice in the frequency domain. Considering the measurement bandwidth of 20 kHz, the measurement capability of our frequency-comb-based spectral interferometer can cover the recording of both male and female voices.

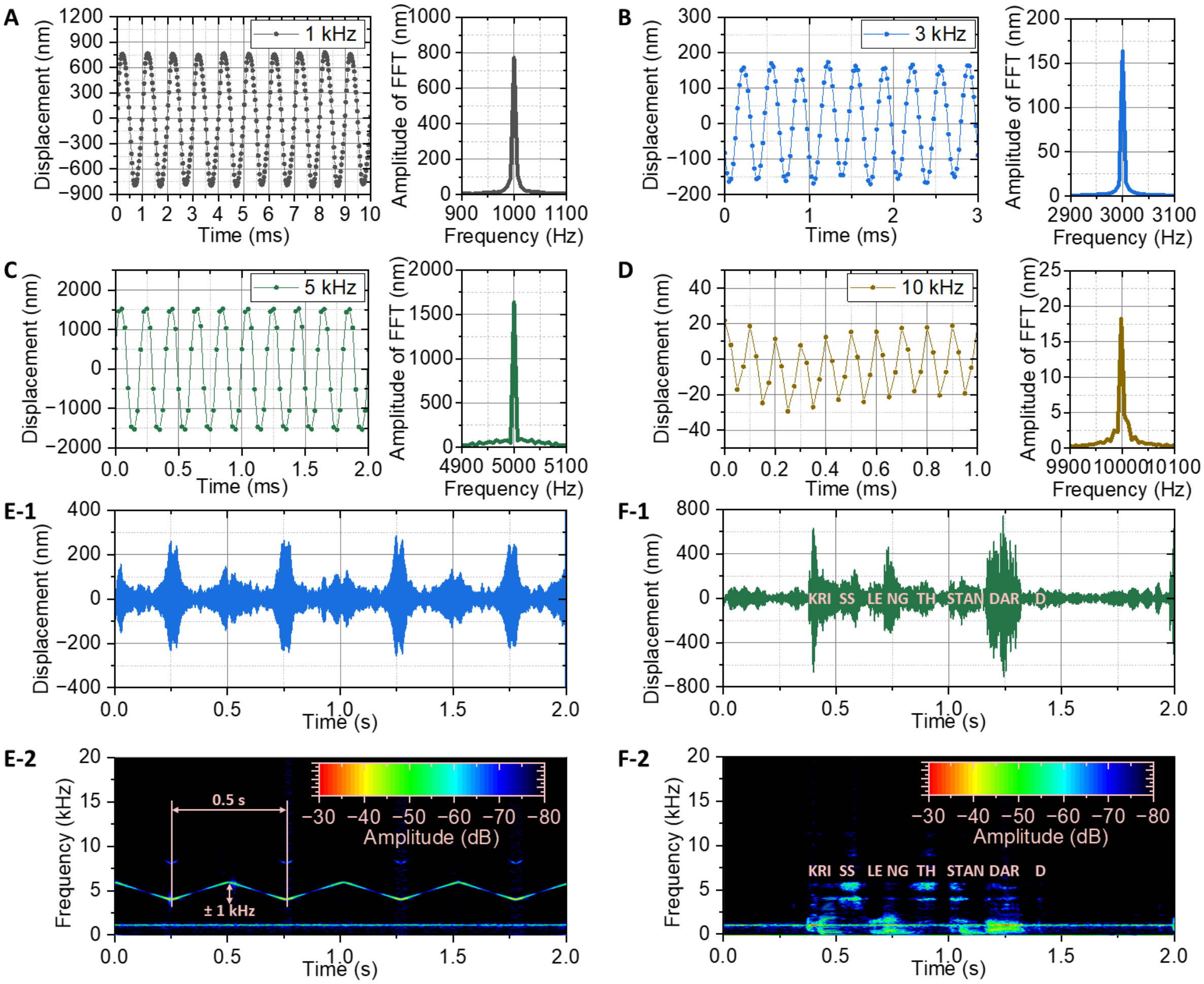

**FIG. 4. Measurement of sound-induced vibrations: A-D**, (Left) Measurement results in the time domain. (Right) Measurement results in the frequency domain, as obtained by Fourier transformation of the time domain results. **E**, Frequency-modulated sound sensing. **E-1**, Measurement results of the frequency-modulated sound in the time domain. **E-2**, Spectrogram of the frequency-modulated sound. **F-1**, Measurement results of a human voice saying "KRISS length standard" in the time domain. **F-2**, Spectrogram of the human voice saying "KRISS length standard." Note that 1.3 kHz is the mechanical resonance frequency of the pellicle BS and that the 1.3 kHz peak in the spectrogram was generated by ambient sound in our laboratory.

## Discussion

Figure 5a summarizes the state-of-the-art distance measurements in terms of the update rate (x-axis) and measurement precision without averaging (y-axis). The diagonal dashed line represents the estimated the white-noise-limited Allan deviation with the function of the averaging time (τ), assuming only white noise with an averaging time of 1 s, ranging from 1 μm to 10 pm. Our measurement method, as indicated by the blue star in Fig. 5A, shows its capabilities, was precision of 0.67 nm without averaging and with an update rate of 40 kHz (25 μs of the measurement time). Fig. 5B summarizes the state-of-the-art distance measurements in terms of the Allan deviation. In a very stable environment such that target movement can be neglected, our technology achieves measurement precision of 3.2 pm with an averaging time of 1 s. There are distance-measurement technologies that are capable of sub-nm precision through sufficient averaging with stable environments (*17*) or that offer faster measurements than our technology (*23,44,45,47*). However, considering both the measurement precision and speed, we believe that the capability of our measurement method makes it worthy of use as the next-generation length standard.

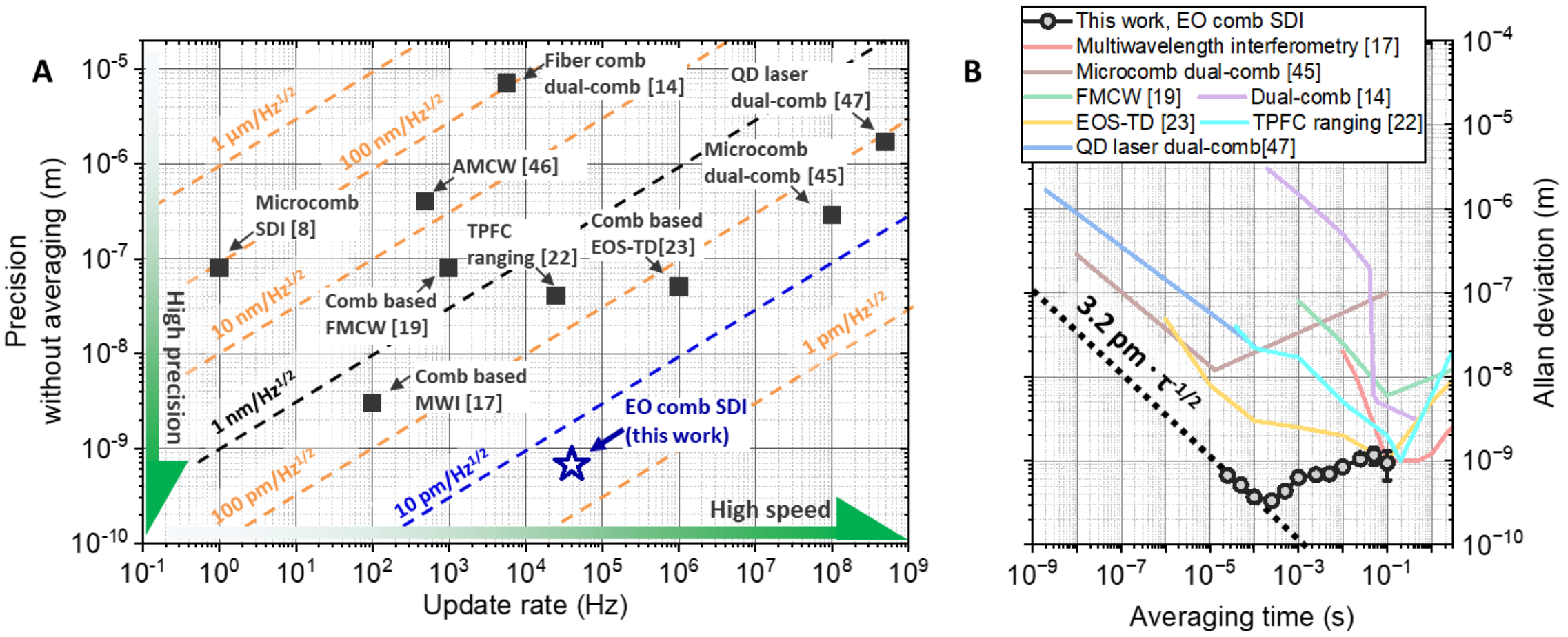

**FIG. 5. Comparison of the measurement capability with the state-of-the-art distance measurement method: A,** Comparison in terms of the measurement precision and speed. The horizontal axis shows the update rate (measurement speed) and the vertical axis shows the precision without averaging. The dashed line shows the measurement sensitivity bounded by white noise. The dotted line shows the estimated white-noise-limited Allan deviation with the function of the averaging time (τ). **B,** Comparison in terms of the Allan deviation. For a comparison with the state-of-the-art distance measurement method, we present the performance capabilities of microcomb-based spectral interferometry (*8*), frequency-comb-based dual-comb ranging (*14*),

multi-wavelength interferometry (*17*), frequency-comb-based FMCW ranging (*19*), time-programmable frequency comb (TPFC) ranging (*22*), frequency-comb-based electro-optic sampling timing detection (EOS-TD) (*23*), microcomb dual-comb ranging (*45*), AMCW ranging (*46*), and a quantum-dash laser dual-comb ranging (*47*).

In this study, we demonstrated distance measurements using a frequency-comb-based spectral interferometry with sub-nm precision and shot-noise limited measurement sensitivity, and we investigated the origins of its measurement precision. We predicted that the fundamental limitation of measurement precision would be the intensity noise, considering that the frequency noise of the frequency comb is such that it can be considered to be negligible. To verify this hypothesis, we compared the measurement precision predicted from a theoretical model with experimental results. The measurement sensitivity was found to be $4.5 \cdot 10^{-12}$m/Hz$^{1/2}$, close to the quantum-limited level (shot-noise-limited), with measurement precision of 0.67 nm (0.34 nm) at an averaging time of 25 μs (250 μs). We confirmed our experimental results were in good agreement with our prediction model considering both the intensity noise and frequency noise of the EO comb. Through this, we experimentally validated our core claim that the intensity noise is the fundamental limit of the measurement precision in frequency-comb-based spectral interferometry and that the frequency noise becomes gradually dominant as the target distance increases.

In conclusion, frequency-comb-based spectral interferometry can provide powerful advantages, including traceability to length standards, ultra-precision, and real-time measurement capabilities as well as long measurement ranges for distance measurements. Our study will be a significant step towards establishing frequency-comb-based spectral interferometry as the next-generation length standard.

## Materials and Methods

**Ranging experiment.** A spectrally flat EO comb with a repetition rate of 18 GHz and spectral bandwidth of 10 THz at the center frequency of 192 THz, which was line-by-line spectrally shaped, was used (*35*). The interferometer is composed of fiber-optic components. A reference beam was generated in the form of a Fresnel reflection beam from the end face of an FC/PC fiber ferrule, which was connected to a fiber collimator. The measurement beam was obtained from a target mirror positioned at about 100 mm. This structure can very aptly define the measurement origin points in distance measurements. The reference and measurement beams were focused into an optical fiber and delivered to a high-speed spectrometer with a measurement speed of 40 kHz through a fiber circulator. The spectral interference between the reference and measurement beams was measured by a high-speed spectrometer and converted into a time-domain signal by Fourier transformation. The time delay ($\tau_{TOF}$) was obtained from peak detection of the time-domain signal. The target distance ($L=v\cdot\tau_{TOF}/2$) was simply converted using the time delay and speed of light in air ($v$).

**Sound sensing experiment.** A pellicle beam splitter (BP145B3, Thorlabs) was positioned at approximately 1 m and a 2 W speaker was installed 100 mm away from the pellicle beam splitter. The speaker was driven by a function generator (DG4162, RIGOL) to generate acoustic waves. For human voice measurements, a recorded sound was played using a mobile phone to minimize contamination by human aerosols.

**Shot-noise calculation.** We assumed that the shot noise in the CCD output voltage is identical for all pixels because the output voltage of each pixel in the high-speed spectrometer was nearly identical for the spectrally flat comb sources. The output voltage of the CCD pixel ($V_{CCD}$) responding to the frequency comb was 370 mV for an update rate of 40 kHz. Considering a CCD conversion efficiency rate of 128 nV/$e^-$, a number of electrons induced by incoming photons ($e_{ccd}$) was about $2.9\cdot10^6$. The number of electron generated by shot noise was approximately 1700 ($e_{shot}=e_{ccd}^{1/2}$), corresponding to the shot-noise-induced output voltage ($P_{shot}$) of 218 μV. The relative intensity noise ($P_{shot}/P_{CCD}$) from the shot noise is 218 μV/370 mV = 0.00059 (0.059 %). The power spectral density of the shot-noise-induced voltage ($S_{\delta Io_shot}(f)$) can be expressed as $S_{\delta Io_shot}(f)=(P_{shot}/P_{CCD})^2/B$, where B is the measurement bandwidth of 20 kHz. The calculated power spectral density of the shot-noise-induced voltage was $S_{\delta Io_shot}(f) = 1.7\cdot10^{-11}$/Hz (or 107.7 dBc/Hz). Based on eq.(3), the power spectral density of the shot noise induced relative intensity noise of spectral interference ($S_{\delta I_shot}(f)$) is $S_{\delta I_shot}(f) = (1+V^2/2)\cdot S_{\delta Io_shot}(f) = 1.18\cdot S_{\delta Io_shot}(f)$ $=2\cdot10^{-11}$/Hz for $V$ of 0.6.

**Calculation of the shot-noise-limited distance sensitivity.** Considering only the intensity noise component of white noise ($S_{white}(f)$) with the same amplitude throughout the spectral range, we simulated the power spectral density of the measured distance ($S_{distance}(f)$, unit: $m^2$/Hz). Applying an interference signal with a known distance to our EO comb spectrum and introducing known white noise, we numerically examined the correlation between $S_{white}(f)$ and $S_{distance}(f)$. The relationship between $S_{white}(f)$ and $S_{distance}(f)$ was found to be $S_{distance}(f) = G\cdot S_{white}(f)$, where G is the conversion efficient in units of $m^2$ (see further details in Appendix G). Through our simulation, the conversion efficient G was found to be $7\cdot10^{-13}$ $m^2$. Consequently, the shot-noise-limited power spectral density of the measured distance ($S_{distance_shot}(f)$) was predicted as $S_{distance_shot}(f) = 1.2\cdot10^{-23} m^2/Hz$. This corresponds to an amplitude spectral density of $3.4\cdot10^{-12} m/Hz^{1/2}$. Note that the actual intensity noise ($\Delta I(f_i,t) = \sqrt{(1+V^2/2)} \cdot \Delta I_o(f_i,t)$) differs with regard to the noise level of $f_i$; therefore, this model is not suitable for actual distance measurements.

**Calculation of the frequency-noise-limited distance sensitivity.** Frequency noise can be converted into intensity noise as described in Eq. (3). The frequency noise of our EO comb is dominated by the distributed feedback laser used as the seed laser. In our previous works (*33*), the effects of frequency variations of the seed laser were found to be negligible, as spectral interferometry uses the period of the spectral interference, not the optical phase. However, due to the slightly different recording times for the spectral interference signal of each comb modes, the frequency noise of the *i*-th frequency comb mode ($\delta_{fi}(t)$) can be considered to be integrated rms frequency noise ($\sigma_{rms}(f_i)$) ranging from 40 kHz (1/spectrometer acquisition time) to 50 kHz (1/CCD exposure time). Through self-homodyne detection (see further details in Appendix H), the integrated rms frequency noise was measured and found to be 110 kHz, acting as random noise. Based on Eq. (3), the output voltage of the frequency-noise-induced intensity noise ($V_{FN}$) can be expressed as $V_{FN}=0.707\cdot(4\pi V_{CCD} VL/v)\cdot \sigma_{rms}(f_i)$. For example, the intensity noise of the CCD output voltage from the frequency noise is 724 μV (72.4 μV) at $L$ = 1,000 mm ($L$=100 mm). Akin to the calculation of the shot-noise-limited distance sensitivity, the frequency-noise-limited distance sensitivity was calculated and determined to be $(L/\mathrm{m})\cdot1.2\cdot10^{-11}$ $m/Hz^{1/2}$, where the unit of $L$ is m.

## Supplementary Materials

Appendix A: Principle of frequency comb based spectral domain interferometry

Appendix B: Data processing for precise peak detection

Appendix C: Broadband EO comb generation

Appendix D: Programmable spectral shaping by a post-data process

Appendix E: Relative intensity noise (RIN) measurement of the EO comb

Appendix F: Effect of electrical power line noise

Appendix G: Relationship between the white-noise-limited PSD of the RIN and the PSD of the distance

**Acknowledgments**: The authors thank Dohyeon Kwon (KRISS) for the discussions about the frequency noise analysis.

**Funding**: This work is supported by the Korea Research Institute of Standards and Science (23011361, 24011043).

**Author contributions:** Y.-S. J. conceived the idea and designed the experiments. Y.-S. J. performed the electro-optic frequency comb generation and distance measurements. S. E. performed the spectral flattening of the EO comb. Y.-S. J. conducted the programmable spectral shaping of the EO comb. H. A. and J. J. developed the high-speed spectrometer. Y.-S. J., J. P. and J. J. conducted the data process of the distance measurement. Y.-S. J., J. P. and J. J. conducted the analysis of the measured data. All authors prepared the manuscript.

**Competing interests:** The authors declare that they have no competing interests.

**Data and materials availability:** All data needed to evaluate the conclusions in the paper are present in the paper and/or the Supplementary Materials. Additional data related to this paper may be requested from the authors.

## Supplementary material

# Approaching the quantum-limited precision in frequency-comb-based spectral interferometric ranging

Yoon-Soo Jang[1,2,*], Heulbi Ahn[3], Sunghoon Eom[1], Jungjae Park[1,2], and Jonghan Jin[1,2]

[1]Length Standard Group, Division of Physical Metrology, Korea Research Institute of Standards and Science (KRISS), 267 Gajeong-ro, Yuseong-gu, Daejeon, 34113, Republic of Korea

[2] Department of Science of Measurement, University of Science and Technology (UST), Daejeon, 34113, Republic of Korea

[3]Meterlab Corp., Daejeon, 34113, Republic of Korea

*ysj@kirss.ac.kr

## 1. Appendix A: Principle of frequency comb based spectral domain interferometry

Figure S1 shows the basic principle of spectral domain interferometry using a frequency comb. In the frequency domain, individual modes of the frequency comb are evenly spaced according to the repetition rate ($f_r$). For example, *i*-th frequency comb modes can be expressed as $f_i = f_{seed} + i \cdot f_r$, where $f_{seed}$ is the frequency of the seed laser and $i$ is a real number ($-i \ldots 0 \ldots i$). In the time domain, the frequency comb can be expressed as a series of evenly spaced pulses with a temporal interval between consecutive pulses ($\tau_{pp} = 1/f_r$). For distance measurements, the time delay ($\tau_{TOF}$) for the round-trip distance at position $L$ can be expressed as $\tau_{TOF} = 2n_{air}L/c$, where $n_{air}$ is the refractive index of air and $c$ is the speed of light. In spectral domain interferometry, $\tau_{TOF}$ can be determined by measuring and analyzing the spectral interference. As shown in Fig. S1, when two pulses (reference pulse and measurement pulse) separated by $\tau_{TOF}$ interact, they produce spectral interference with a period of $1/\tau_{TOF}$. This can be expressed as

$$I(f_i, L) = I_{ref}(f_i) + I_{mea}(f_i) + 2V\sqrt{I_{ref}(f_i) \cdot I_{mea}(f_i)}\cos(2\pi f_i L/c), \qquad \text{(S1)}$$

where $I_{ref}$ is the intensity of the reference beam, $I_{mea}$ is the intensity of the measurement beam, and $V$ denotes the visibility. Equation (1) in the main text is a simplified version of this. The measured spectral interference is then converted into the time-domain signal through Fourier transformation. The target distance $L$ can be determined by measuring $\tau_{TOF}$ from the peak position of the reconstructed time-domain signal.

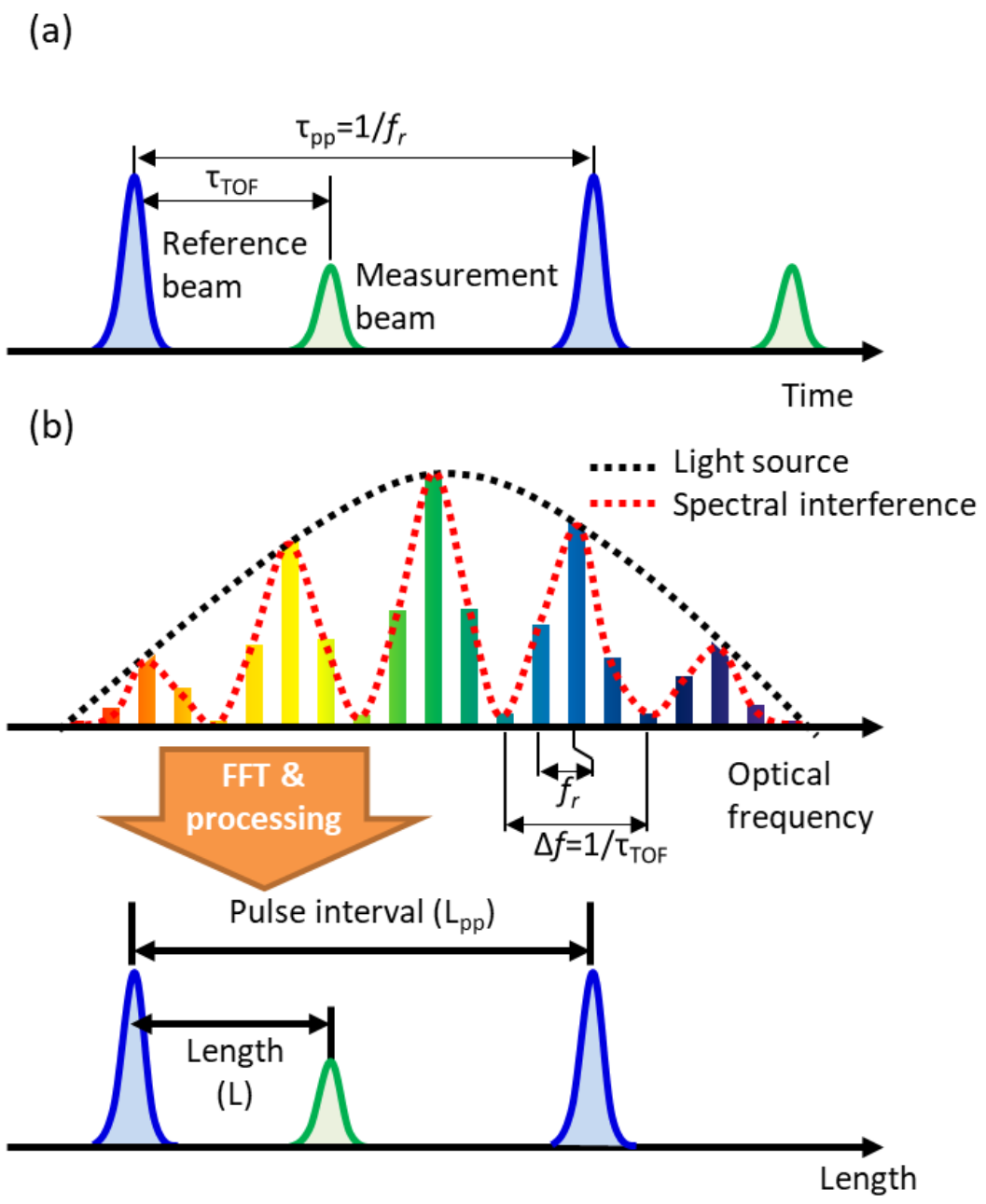

**Fig. S1. Principle of spectral interferometry:** (a) Separation between the reference and measurement pulse in the time domain. (b) Spectral interference and reconstructed length-domain signal obtained from Fourier transformation of the spectral interference.

## 2. Appendix B: Data processing for precise peak detection

For high-precision determination of the peak position $\tau_{TOF}$, many methods have been proposed, including polynomial fitting, the use of a centroid algorithm, and a phase slope method. Compared to simply selecting the amplitude peak of the Fourier transform, these methods enable sub-pixel precision so as to achieve nanometric measurement precision. In our case, polynomial fitting was used. The peak signal in the reconstructed time domain can be modeled as $I(\tau) = A\tau^2 + B\tau + C$. The peak position is determined at the point where the first derivative ($dI(\tau)/d\tau = 2A\tau + B$) is zero. Then, the peak position can be determined as $\tau = -B/2A$.

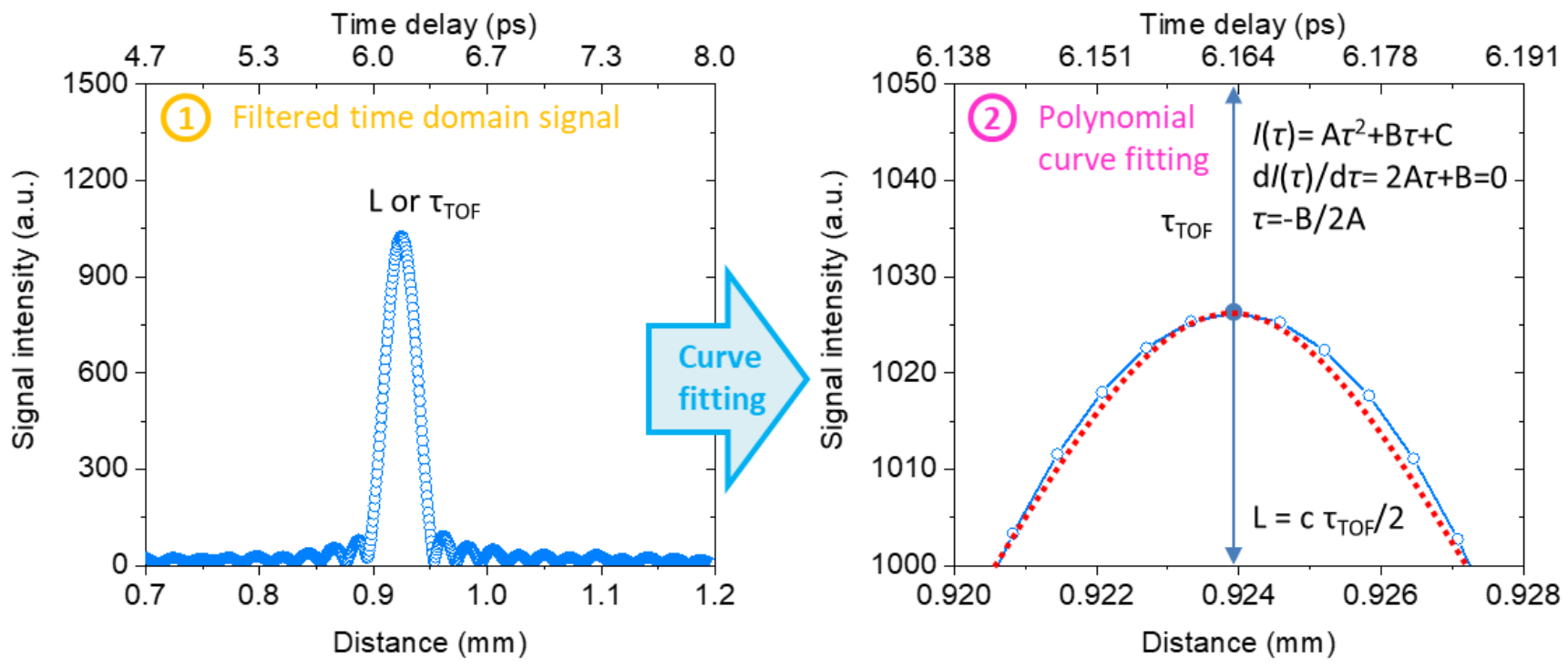

**Fig. S2. Data processing of precise peak detection by polynomial curve fitting**

### 3. Appendix C: Broadband EO comb generation

Figure S3 shows the optical layout and optical spectrum of the spectrally broad and flat EO comb. It consists of three parts. The first of these is for seed EO comb generation. The seed EO comb was generated using a typical method involving a CW laser, an intensity modulator and three phase modulators (*48*). All modulators were driven by single RF synthesizer, which was locked to a GPS-based frequency reference with uncertainty of $10^{-13}$. The seed EO comb typically has a spectral bandwidth of <1 THz, which is not enough to support high-precision ranging. The second part handles super-continuum generation. To determine the broadband spectrum for high-precision ranging, the seed EO comb was temporally compressed by 300 m of single-mode fiber and amplified to 2 W by an erbium-ytterbium co-doped fiber amplifier and then positioned incident to 100 m of high-nonlinear fiber for supercontinuum generation. The spectral broaden EO comb typically has a spectral bandwidth of >10 THz. The third part involves line-by-line spectral shaping. To generate the spectrally flat EO comb, a pulse shaper is used for line-by-line spectral shaping. Figs. S3 b-d correspondingly show the optical spectrum of the seed EO comb, the spectral broaden EO comb and the spectrally flatted EO comb.

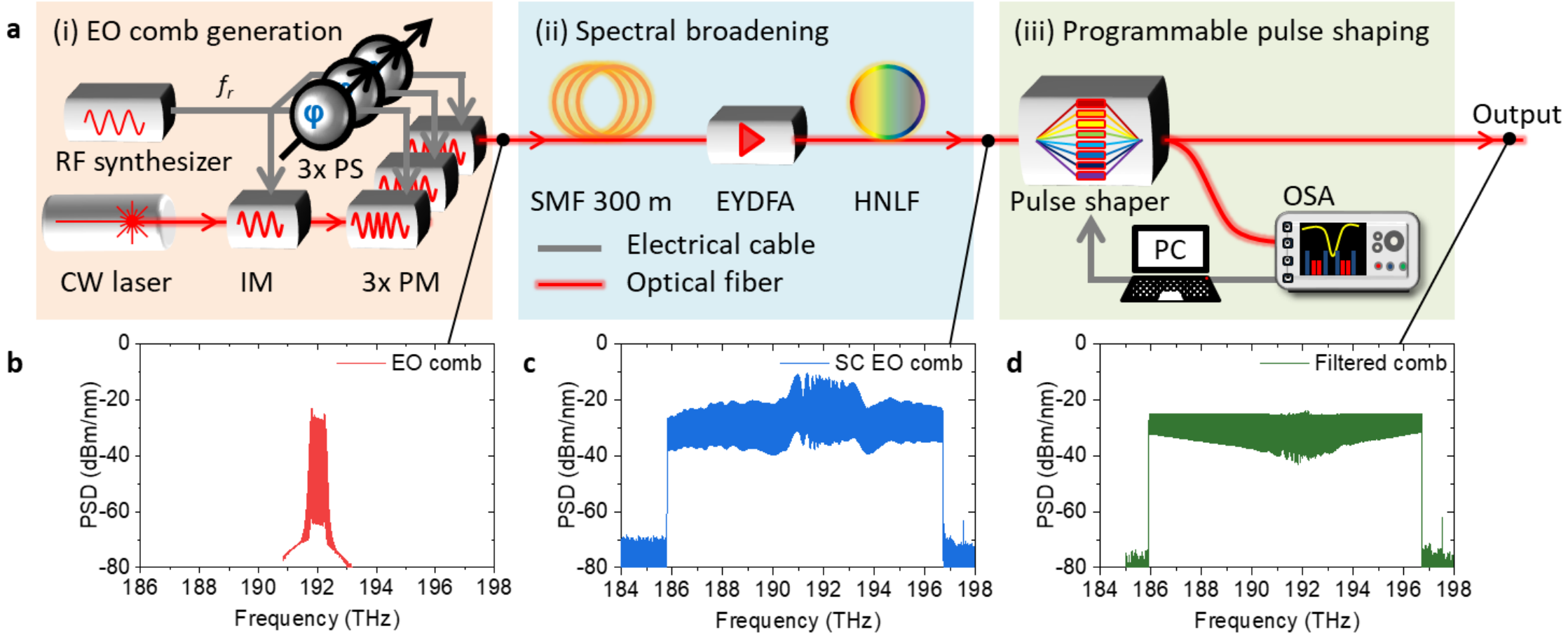

**Fig. S3. Spectrally flat and broadband EO comb generation:** a, Optical layout for spectrally flat EO comb generation. b, Seed EO comb spectrum. c, Spectral broaden EO comb spectrum. d, Spectrally flat EO comb by programmable line-by-line comb shaping.

## 4. Appendix D: Programmable spectral shaping by a post-data process

Figure S4 shows programmable spectral shaping by a post-data process. To generate an arbitrary spectral shape, a square shape is preferred. In the spectrometer, the spectral flatness of the spectrally shaped EO comb was not sufficiently uniform due to limitations of the pulse shaper and the spectrometer. To generate the square spectral shape, the EO comb spectrum ($I_o(f_i)$) was measured and used to normalize the spectral interference ($I(f_i)$). As shown in Fig. S4, the spectral interference was normalized by the EO comb spectrum. In a previous study by the authors, we revealed that super-Gaussian shaped spectral interference provides more high-precision distance measurements (*41*). We applied a 10-th super-Gaussian window with a spectral bandwidth of 9 THz.

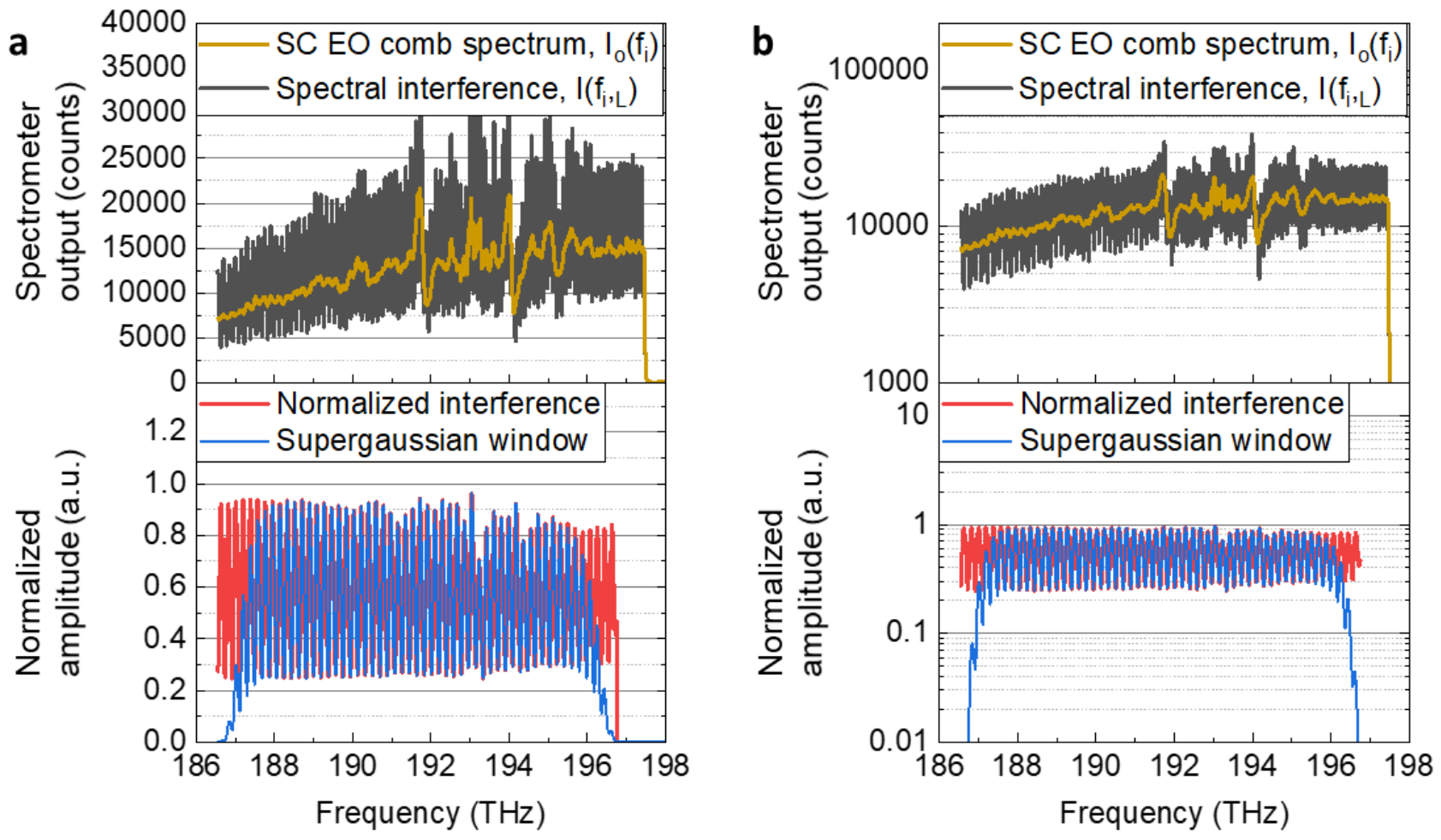

**Fig. S4. Spectral shaping through post-data processing:** a, Spectral shaping through post-data processing in linear scale. b, Spectral shaping through post-data processing in log scale.

## 5. Appendix E: Relative intensity noise (RIN) measurement of the EO comb

Figure S5 shows the RIN measurement of the square comb as measured by a high-speed spectrometer. As shown in Fig. S5a, RIN levels differ in different frequency ranges. For the lower RIN range, the power spectral density of the RIN reaches the shot noise limit of $1.7 \cdot 10^{-11}$/Hz (or 107.7 dBc/Hz). However, certain regions, highlighted in red in Fig. S5a, exhibit power spectral density levels that exceed the shot noise limit. Fig. S5b shows power spectral density examples for low-noise and high-noise states. Fig. S5c shows examples of the relative intensity noise ($\Delta I_o/I_o$) of light sources at 194.8 THz (high-noise region) and 189.2 THz (low noise region) in the time domain. This measured intensity noise was factored into a simulation of the effects of the intensity noise on the measurement precision.

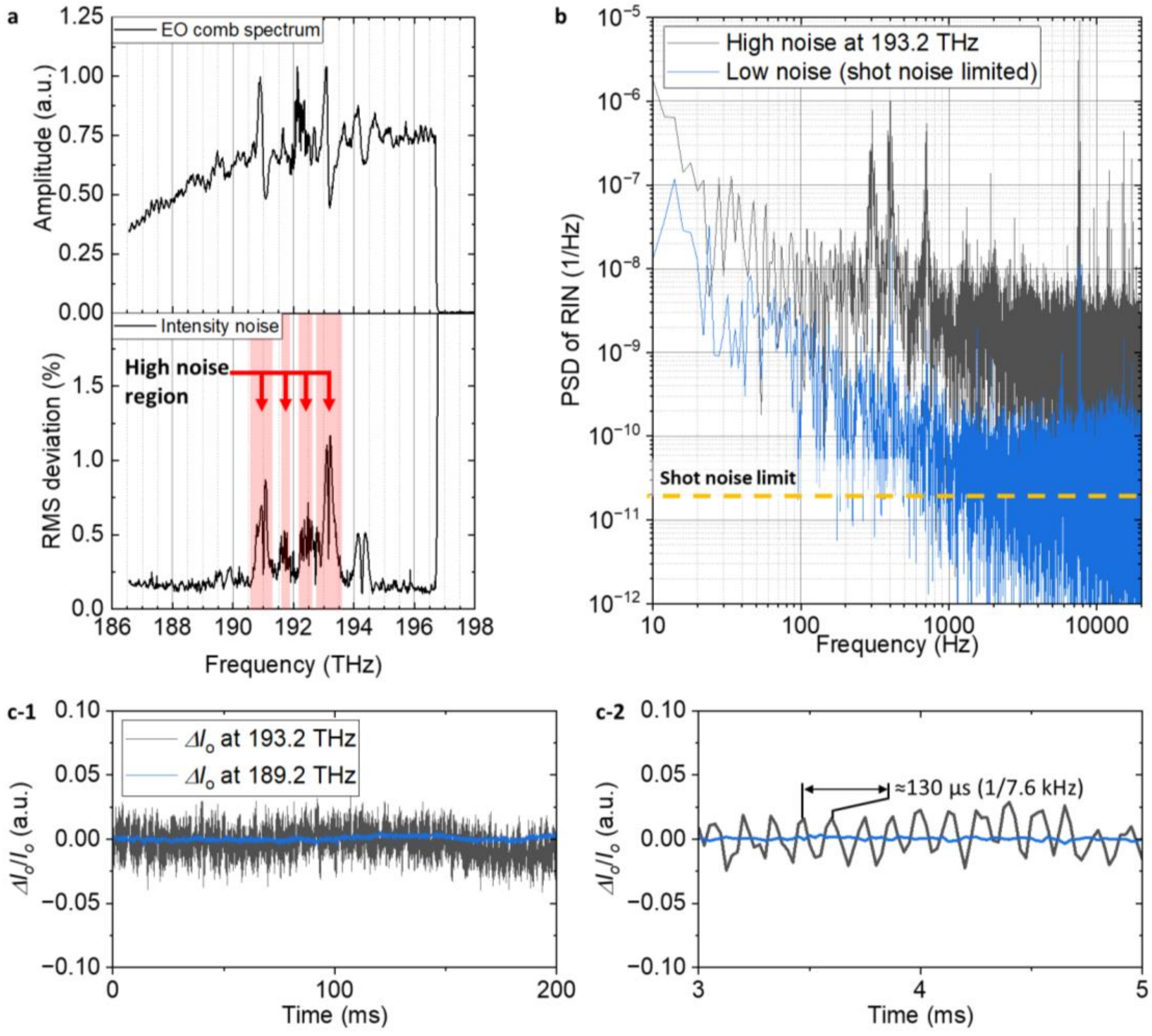

**Fig. S5. RIN measurement using high-speed spectrometer: a,** Spectrally flat broadband EO comb spectrum and corresponding RMS amplitude deviation. **b,** Power spectral density of the relative intensity noise in the high-noise and low-noise states. **c,** Examples of relative intensity

noise ($\Delta I_o/I_o$) of the light sources at 194.8 THz (high-noise region) and 189.2 THz (low-noise region).

Figure S6 shows further details about the RIN. Fig. S6a shows the RIN measured by the photodiode and FFT analyzer and the RIN contribution from each component. 1/*f* noise occurred in the 2 W EDFA, and peak at 7.8 kHz was generated from the seed EO comb, which can be considered as electrical noise. Fig. S6b compares the power spectral density of the RIN measured by the photodiode and the high-speed spectrometer. By comparing these results, the intensity noise levels of the light source and the spectrometer can be examined. The 1/f noise appears higher in the spectrometer measurement, indicating a limitation of the spectrometer. This could be improved by reducing thermal noise through the use of a low-temperature spectrometer. The RIN observed through the photodiode is lower than the CCD's shot noise; however, the white noise obtained through the spectrometer is slightly higher than the shot noise. This occurs because the measurement limit for each pixel of the spectrometer is shot noise, and RIN higher than the CCD shot noise was observed in a specific wavelength range. Therefore, the white noise level obtained through the spectrometer appears higher than the shot noise.

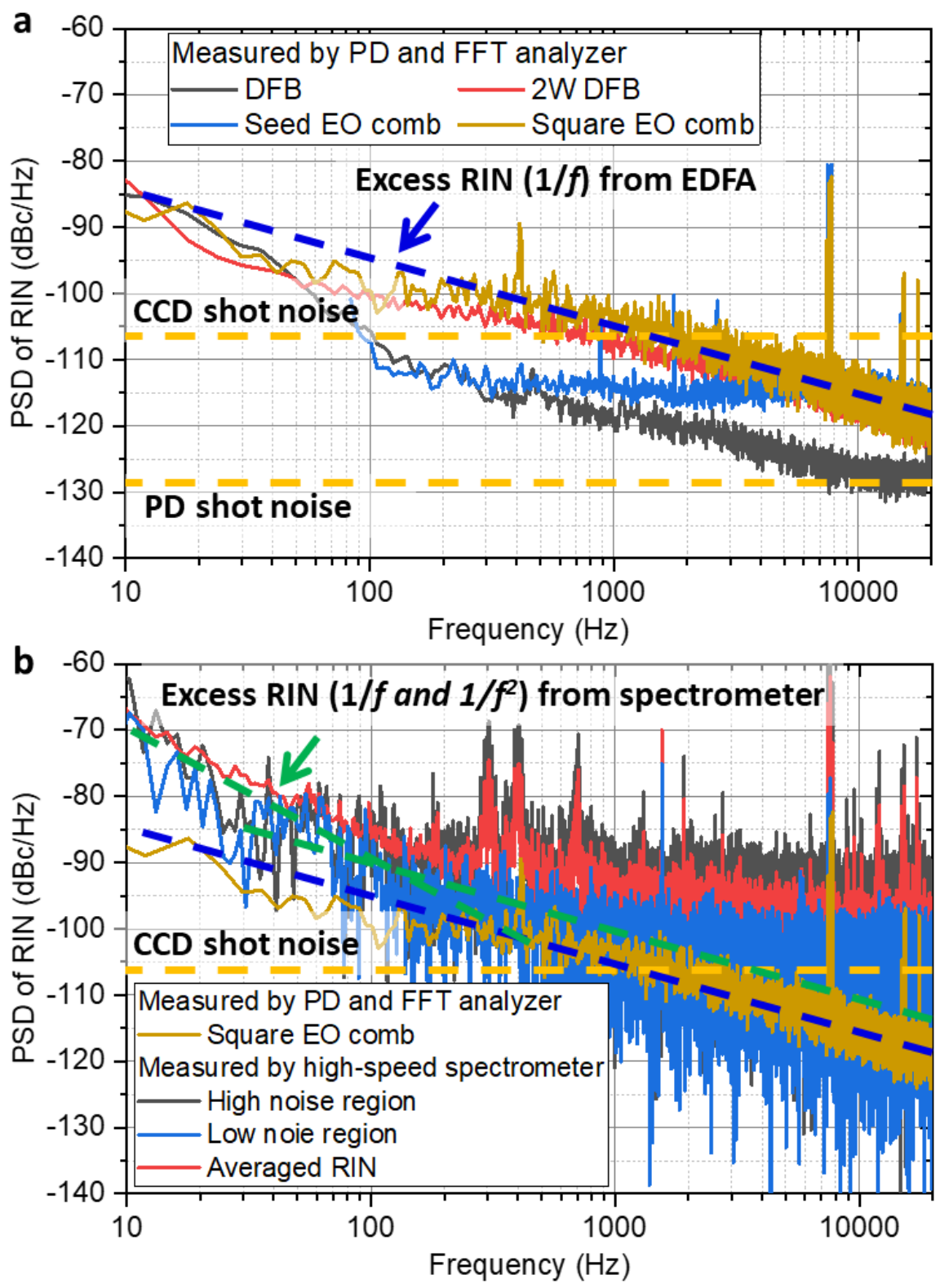

**Fig. S6. RIN comparison from the photodiode and the high-speed spectrometer: a,** RIN contribution from each component. **b,** Comparison of the power spectral density of RIN measured by the photodiode and the high-speed spectrometer.

## 6. Appendix F: Effect of electrical power line noise

Figure S7 explains the electrical power line noise. Peaks were observed in the PSD of the length measurement around 100 Hz to 400 Hz and near 7.5 kHz. To investigate the cause of these peaks, the output of the power supply was observed. Peaks in the same frequency range were observed even from the DC power supply connected throughout the EO comb system. Therefore, it can be considered that these peak components are induced from power supply noise. To improve this, measures such as installing EMI filters to reduce laboratory power noise can be taken.

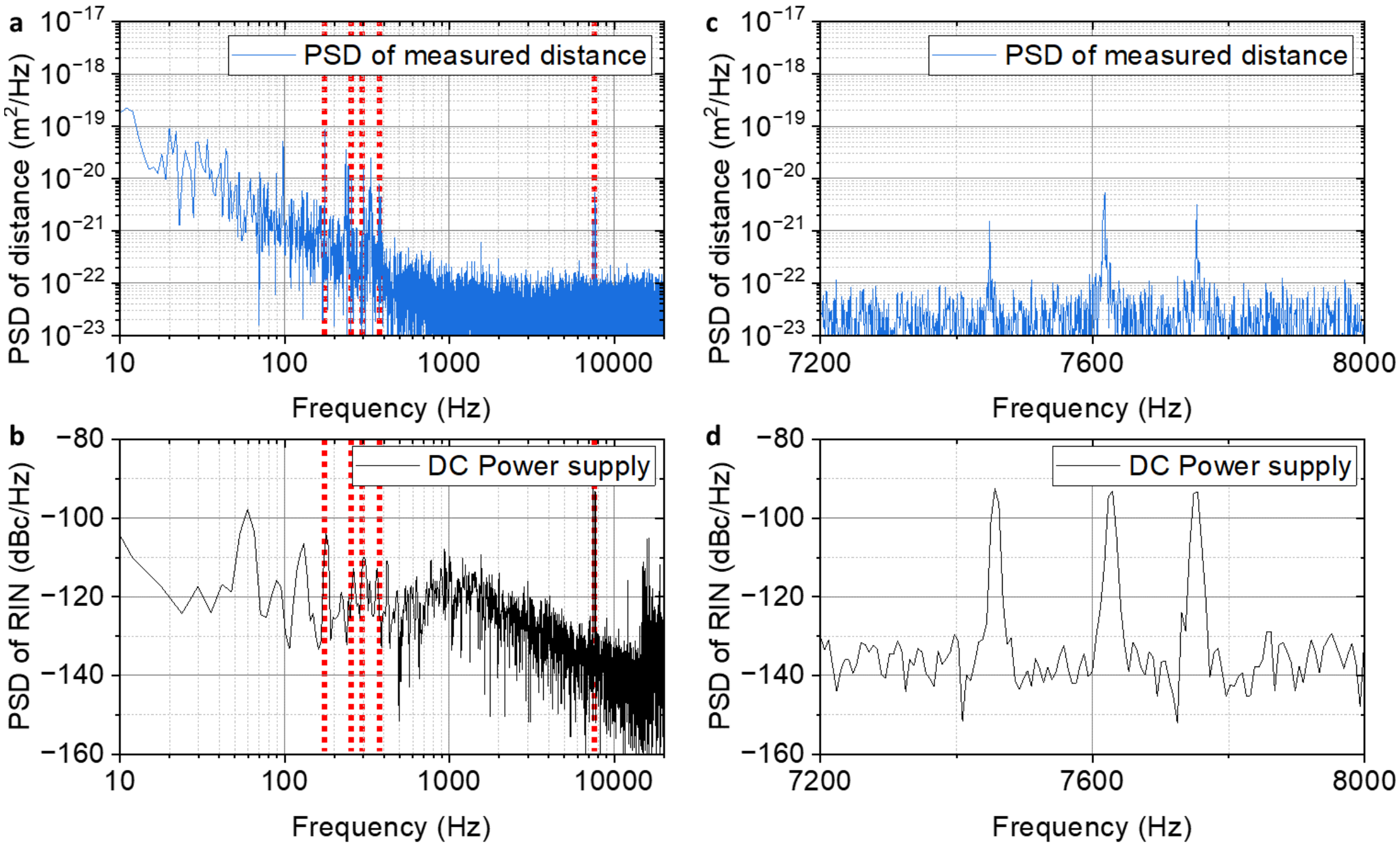

**Fig. S7. Comparison between the power spectral density of the measured distance and the DC power supply: a,** PSD of the measured distance. **b,** PSD of the RIN for the DC power supply. **c,** magnified view for the PSD of the measured distance. **d,** magnified view of the PSD of the RIN for the DC power supply.

## 7. Appendix G: Relationship between the white-noise-limited PSD of the RIN and the PSD of the distance

Figure S8 presents the results of a simulation conducted to examine the impact on distance-measurement results when there is only white noise with an amplitude identical to that of the RIN ($S_{white}(f)$). By injecting random noise (Fig. S8a) into the ideal interference signal with a known distance and visibility of 0.6, the power spectral density of the virtually measured distance ($S_{distance_white}(f)$) can be calculated, as shown in Fig. S8b. Considering the model equation $S_{distance_white}(f) = G \cdot S_{white}(f)$, by taking the moving average of the PSD of RIN and the PSD of the distance to determine the G value, results similar to Fig. S8c were obtained. Through the simulation, we confirmed that the constant G has a value of $7 \cdot 10^{-13}$ m$^2$.

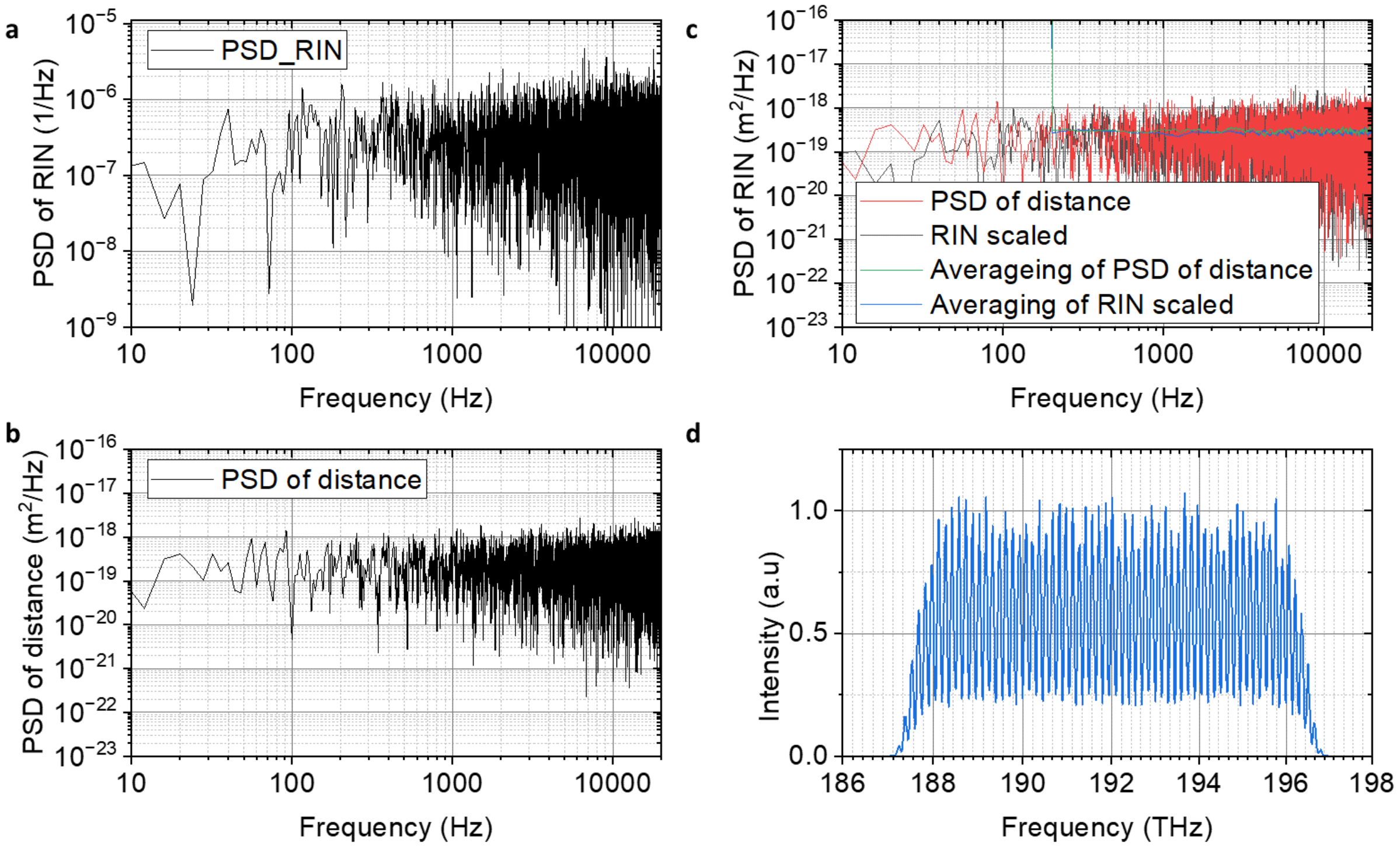

**Fig. S8. Simulation of the white-noise-limited power spectral density of the measured distance: a,** PSD of virtual RIN. **b,** PSD of the virtually measured distance. **c,** Comparison between the PSD of the virtual RIN and virtually measured distance to determine the relationship. **d,** Simulated spectral interference with visibility of 0.6.

## 8. Appendix H: Frequency noise measurement by self-homodyne detection

Figure S9 shows the optical layout used during the self-homodyne measurements of the laser frequency noise (*49*). During the self-homodyne measurements, the output voltage ($V_{PD}$) of the balanced photodetector (BPD) can be expressed as $V_{PD} = V_o \sin(\varphi)$, where $V_o$ is the amplitude of the output voltage and $\varphi$ is the phase delay. The phase delay can be expressed as $\varphi = 2\pi \cdot \tau_{delay} \cdot f_{LUT}$, where $\tau_{delay}$ is the time delay between two arms and $f_{LUT}$ is the frequency of the laser under test. Assuming the $\tau_{delay}$ is constant, the deviation of the output voltage ($\Delta V_{PD}$) can be expressed as $\Delta V_{PD} = 2\pi \cdot \tau_{delay} \cdot V_o \cdot \Delta f_{LUT}$ at the approximate point where $\varphi = 0$. Consequently, the frequency noise ($\Delta f_{LUT}$) can be measured according to the relationship of $\Delta f_{LUT} = \Delta V_{PD}/(2\pi \cdot \tau_{delay} \cdot V_o)$. The voltage noise ($\Delta V_{PD}$) was measured using a high-speed oscilloscope.

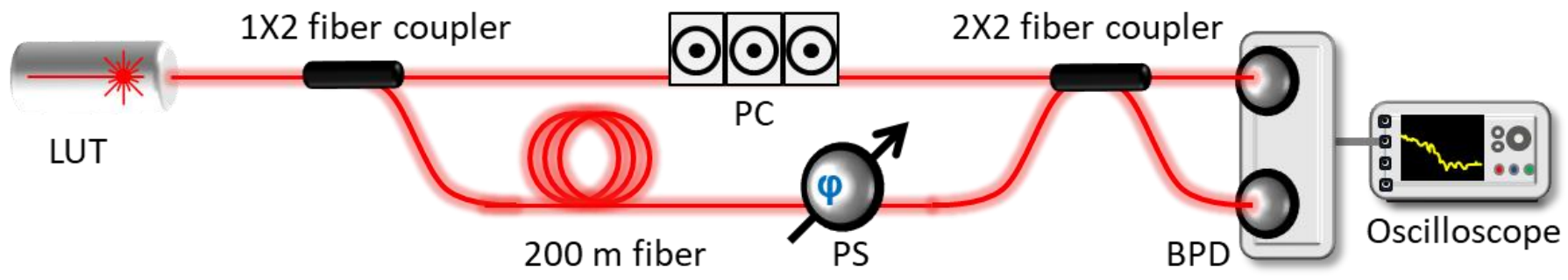

**Fig. S9. Optical layout of the self-homodyne measurement of measure the laser frequency noise.** LUT: laser under test, PC: polarization controller, PS: phase shifter, BPD: balanced photodiode

Note that the frequency noise of the spectrally broad and flat EO comb was dominated by the DFB laser used as a seed laser with a few MHz linewidth, not by electrical devices such as RF synthesizer, phase modulator and RF drivers. [S4] The time-dependent frequency variation of the DFB laser was found to be 4 MHz with a sampling time of 770 ns. As shown in Fig.S10 b, power spectral density of frequency noise was dominated by white noise beyond 20 kHz. Below 20 kHz, power spectral density of frequency noise was dominated by flicker noise and random walk. The effective frequency noise ($\Delta\delta_{fi}(t)$) that contributes to fluctuations in distance measurements is the frequency stability at single measurement time of high-speed spectrometer (sampling time, 25 μs) [S5]. The frequency noise ($\Delta\delta_{fi}(t)$) was calculated from the power spectral density of frequency noise using the Enrico chart and it was found to be 133 kHz at 25 μs. [42]

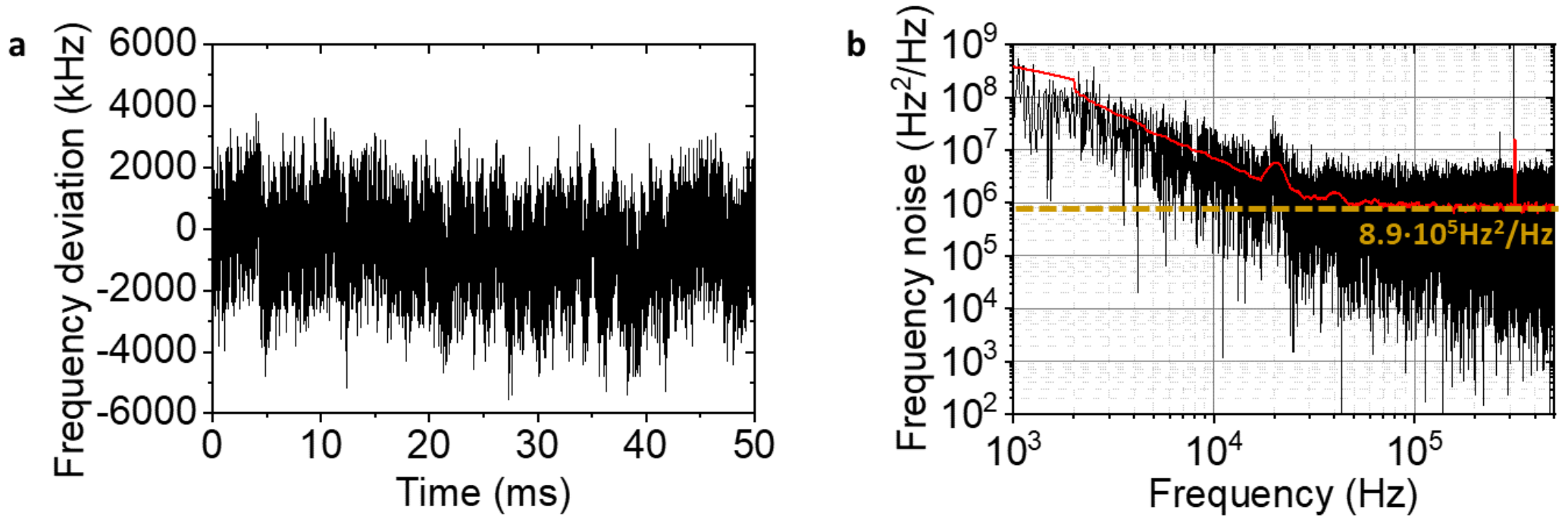

**Fig. S10. Frequency noise measurement. a**, Time-dependent frequency variation of the DFB laser. **b**, Power spectral density of the frequency noise from 1 kHz to 50 kHz on the log scale.

## 9. Appendix I: Measurement performance comparison with the-state-of-the-art absolute distance-measurement methods

Table S1 and S2 present a performance comparison with state-of-the-art absolute distance-measurement methods and earlier spectral interferometry methods, respectively.

**Table S1. Summary Comparison between State-of-the-Art Absolute Distance Measurement Methods and the EO comb Spectral Interferometry**

| Ref. | Metrology approach | Comb/Mod. rep. rate | Non-ambiguity range | White-noise-limited precision (1 σ) | Precision (1 σ) | Sampling rate |
|---|---|---|---|---|---|---|
| [14] | Fiber-comb-based dual-comb interferometry | 100.021 MHz & 100.016 MHz | 1.5 m | 40 nm·$\tau^{-1/2}$ (Peak detection) 1 nm·$\tau^{-1/2}$ (Interferometric method) | 3 nm @ $\tau_{avg}$=0.5 s | 5 kHz |
| [17] | Frequency-comb referenced multi-wavelength interferometry (MWI) | 100 MHz | 3.8 m | 2 nm·$\tau^{-1/2}$ | 0.57 nm @ $\tau_{avg}$=100 s | 100 Hz |
| [19] | Frequency-comb-assisted frequency-modulated continuous wave (FMCW) | 1 THz | N/A | 2 nm·$\tau^{-1/2}$ | 6 nm @ $\tau_{avg}$=0.1 s | 1 kHz |
| [22] | Time-programmable frequency comb ranging | 200 MHz | 750 mm | 200 pm·$\tau^{-1/2}$ | 1 nm @ $\tau_{avg}$=0.2 s | 8.3 kHz |
| [23] | Frequency-comb-based electro-optic sampling timing detection | 250 MHz | 300 mm | 50 pm·$\tau^{-1/2}$ | 1 nm @ $\tau_{avg}$=0.1 s | 1 MHz |
| [33] | EO comb SDI (our previous work) | 17.5 GHz | 4.3 mm | 10 nm·$\tau^{-1/2}$ | 50 nm @ $\tau_{avg}$=67 ms | 3 kHz |
| [45] | Microcomb dual-comb | 95.7 GHz & 95.8 GHz | 1.6 mm | 29 pm·$\tau^{-1/2}$ | 12 nm @ $\tau_{avg}$=13 µs | 96.4 MHz |
| [46] | Amplitude-modulation continuous wavelength (AMCW) | 15 GHz | 10 mm | 22 nm·$\tau^{-1/2}$ | 43 nm @ $\tau_{avg}$=0.4 s | 488 Hz |
| [47] | Integrated comb ranging (dual-comb) | 49.7 GHz & 50.2 GHz | 3 mm | 100 pm·$\tau^{-1/2}$ | 23 nm @ $\tau_{avg}$=101 µs | 495 MHz |
| **This work** | **EO comb spectral interferometry** | **18 GHz** | **4.2 mm** | **3.2 pm·$\tau^{-1/2}$** | **0.33 nm @ $\tau_{avg}$=250 µs** | **40 kHz** |

**Table S2. Summary Comparison between Spectral Interferometry and EO Comb Spectral Interferometry**

| Ref. | Metrology approach | Comb/Mod. rep. rate | Non-ambiguity range | White-noise-limited precision (1 σ) | Precision (1 σ) | Sampling rate |
|---|---|---|---|---|---|---|
| [8] | Solid-state frequency-comb-based SDI | 1 GHz | 150 mm | N/A | N/A | N/A |
| [25] | Chip-scale soliton-microcomb-based SDI | 88.5 GHz | 1.7 mm | 80 nm·$\tau^{-1/2}$ | 12 nm | 1 Hz |
| [30] | Fiber-frequency-comb-based SDI | 250 MHz | 600 mm | N/A | N/A | 1 Hz |
| [31] | Chip-scale soliton-microcomb-based SDI | 48.9 GHz | 3 mm | 50 nm·$\tau^{-1/2}$ (at 1.2 km) | 27 nm (at 1.2 km) | 35 kHz |
| [33] | EO comb SDI (our previous work) | 17.5 GHz | 4.3 mm | 10 nm·$\tau^{-1/2}$ | 50 nm @ $\tau_{avg}$=67 ms | 3 kHz |
| [35] | EO comb SDI (our previous work) | 17.5 GHz | 4.3 mm | NA | 6 nm @ $\tau_{avg}$=25 μs | 40 kHz |
| **This work** | **EO comb spectral interferometry** | **18 GHz** | **4.2 mm** | **3.2 pm·$\tau^{-1/2}$** | **0.33 nm @ $\tau_{avg}$=250 μs** | **40 kHz** |

## Supplementary references